 \documentclass[final,5p,times,twocolumn]{elsarticle}

 \usepackage{graphicx}
 \usepackage{subfigure}
 \usepackage{booktabs}
\usepackage{verbatim}
\usepackage{multirow}
\usepackage{scrextend}
\usepackage[hyphens]{url}
\usepackage[british]{babel}
\usepackage[setpagesize=false, colorlinks=true,  urlcolor=blue]{hyperref}

\usepackage{amssymb}

\usepackage[switch]{lineno}

\journal{}

\begin{document}

\def\nuc#1#2{${}^{#1}$#2}
\def\BBz{0$\nu\beta\beta$}
\def\BBt{2$\nu\beta\beta$}
\def\BB{$\beta\beta$}
\def\Tz{$T^{0\nu}_{1/2}$}
\def\Tt{$T^{2\nu}_{1/2}$}
\def\mj{M{\sc ajo\-ra\-na}}
\def\dem{D{\sc e\-mon\-strat\-or}}
\def\mg{M{\sc a}G{\sc e}}
\def\QBB{Q$_{\beta\beta}$}
\def\mBB{$\left < \mbox{m}_{\beta\beta} \right >$}
\def\ge{$^{76}$Ge}

\begin{frontmatter}

\title{High voltage testing for the \textsc{Majorana Demonstrator}}

\author[lbnl]{N.~Abgrall}		
\author[pnnl]{I.J.~Arnquist}
\author[usc,ornl]{F.T.~Avignone~III}
\author[ITEP]{A.S.~Barabash}	
\author[ornl]{F.E.~Bertrand}
\author[lbnl]{A.W.~Bradley}
\author[JINR]{V.~Brudanin}
\author[duke,tunl]{M.~Busch}	
\author[uw]{M.~Buuck}
\author[usd]{D.~Byram}
\author[sdsmt]{A.S.~Caldwell}
\author[lbnl]{Y-D.~Chan}
\author[sdsmt]{C.D.~Christofferson}
\author[lanl]{P.-H.~Chu}
\author[uw]{C. Cuesta\corref{ca}}
\ead{ccuesta@uw.edu}
\cortext[ca]{Corresponding author}

\author[uw]{J.A.~Detwiler}
\author[uw]{P.J.~Doe}
\author[sdsmt]{C.~Dunagan}
\author[ut]{Yu.~Efremenko}
\author[ou]{H.~Ejiri}
\author[lanl]{S.R.~Elliott}
\author[uw]{Z.~Fu}
\author[ornl]{A.~Galindo-Uribarri}	
\author[unc,tunl]{G.K.~Giovanetti}
\author[lanl]{J.~Goett}	
\author[ornl]{M.P.~Green}
\author[uw]{J.~Gruszko}
\author[uw]{I.S.~Guinn}		
\author[usc]{V.E.~Guiseppe}	
\author[unc,tunl]{R.~Henning}
\author[pnnl]{E.W.~Hoppe}
\author[sdsmt]{S.~Howard}
\author[unc,tunl]{M.A.~Howe}
\author[usd]{B.R.~Jasinski}
\author[blhill]{K.J.~Keeter}
\author[ttu]{M.F.~Kidd}	
\author[ITEP]{S.I.~Konovalov}
\author[pnnl]{R.T.~Kouzes}
\author[pnnl]{B.D.~LaFerriere}
\author[uw]{J.~Leon}	
\author[uw]{A.~Li}
\author[unc,tunl]{J.~MacMullin}
\author[queens]{R.D.~Martin}
\author[lanl]{R. Massarczyk}
\author[unc,tunl]{S.J.~Meijer}	
\author[lbnl]{S.~Mertens}		
\author[pnnl]{J.L.~Orrell}
\author[unc,tunl]{C.~O'Shaughnessy}	
\author[lbnl]{A.W.P.~Poon}
\author[ornl]{D.C.~Radford}
\author[unc,tunl]{J.~Rager}	
\author[lanl]{K.~Rielage}
\author[uw]{R.G.H.~Robertson}
\author[ut,ornl]{E.~Romero-Romero}
\author[unc,tunl]{B.~Shanks}	
\author[JINR]{M.~Shirchenko}
\author[usd]{N.~Snyder}	
\author[sdsmt]{A.M.~Suriano}
\author[usc]{D.~Tedeschi}
\author[uw]{A.~Thompson}
\author[uw]{K.T.~Ton}
\author[unc,tunl]{J.E.~Trimble}
\author[ornl]{R.L.~Varner}
\author[JINR]{S.~Vasilyev}
\author[lbnl]{K.~Vetter\fnref{ucb}}
\author[unc,tunl]{K.~Vorren}
\author[lanl]{B.R.~White}	
\author[unc,tunl,ornl]{J.F.~Wilkerson}
\author[usc]{C.~Wiseman}		
\author[lanl]{W.~Xu}
\author[JINR]{E.~Yakushev}
\author[ornl]{C.-H.~Yu}
\author[ITEP]{V.~Yumatov}

\address[uw]{Center for Experimental Nuclear Physics and Astrophysics, and Department of Physics, University of Washington, Seattle, WA, USA}
\address[lbnl]{Nuclear Science Division, Lawrence Berkeley National Laboratory, Berkeley, CA, USA}
\address[pnnl]{Pacific Northwest National Laboratory, Richland, WA, USA}
\address[usc]{Department of Physics and Astronomy, University of South Carolina, Columbia, SC, USA}
\address[ornl]{Oak Ridge National Laboratory, Oak Ridge, TN, USA}
\address[ITEP]{National Research Center ``Kurchatov Institute'' Institute for Theoretical and Experimental Physics, Moscow, Russia}
\address[lanl]{Los Alamos National Laboratory, Los Alamos, NM, USA}
\address[JINR]{Joint Institute for Nuclear Research, Dubna, Russia}
\address[duke]{Department of Physics, Duke University, Durham, NC, USA}
\address[tunl]{Triangle Universities Nuclear Laboratory, Durham, NC, USA}
\address[usd]{Department of Physics, University of South Dakota, Vermillion, SD, USA}
\address[sdsmt]{South Dakota School of Mines and Technology, Rapid City, SD, USA}
\address[ut]{Department of Physics and Astronomy, University of Tennessee, Knoxville, TN, USA}
\address[ou]{Research Center for Nuclear Physics and Department of Physics, Osaka University, Ibaraki, Osaka, Japan}
\address[unc]{Department of Physics and Astronomy, University of North Carolina, Chapel Hill, NC, USA}
\address[blhill]{Department of Physics, Black Hills State University, Spearfish, SD, USA}
\address[ttu]{Tennessee Tech University, Cookeville, TN, USA}
\address[queens]{Department of Physics, Engineering Physics and Astronomy, Queen's University, Kingston, ON, Canada}
\fntext[ucb]{Alternate Address: Department of Nuclear Engineering, University of California, Berkeley, CA, USA}

\begin{abstract}
The~\mj~Collaboration is constructing the~\mj\ \dem, an ultra-low background, 44-kg modular high-purity Ge (HPGe) detector array to search for neutrinoless double-beta decay in \ge. The phenomenon of surface micro-discharge induced by high-voltage has been studied in the context of the \mj\ \dem. This effect can damage the front-end electronics or mimic detector signals. To ensure the correct performance, every high-voltage cable and feedthrough must be capable of supplying HPGe detector operating voltages as high as 5~kV without exhibiting discharge. R\&D measurements were carried out to understand the testing system and determine the optimum design configuration of the high-voltage path, including different improvements of the cable layout and feedthrough flange model selection. Every cable and feedthrough to be used at the~\mj\ \dem\ was characterized and the micro-discharge effects during the \mj\ \dem\ commissioning phase were studied. A stable configuration has been achieved, and the cables and connectors can supply HPGe detector operating voltages without exhibiting discharge.

\end{abstract}

\begin{keyword}
high-voltage \sep micro-discharge \sep vacuum \sep \mj


\end{keyword}

\end{frontmatter}



\section{Introduction}
\label{sec1}

Neutrinoless double-beta (\BBz) decay is a model-independent method to search for lepton number violation and to determine the Dirac or Majorana nature of the neutrino~\cite{Zralek,Camilleri,Avignone,vergados}. Observation of this rare process would have significant implications for our understanding of the nature of neutrinos and matter in general. The~\mj~\dem~\cite{mjd} is an array of enriched and natural germanium detectors that will search for the \BBz-decay of \ge. The specific goals of the~\mj~\dem~are: to demonstrate a path forward to achieving a background rate at or below 1~count/(ROI-t-y) in the 4-keV region-of-interest (ROI) around the 2039-keV~\QBB~of the \ge\ \BBz-decay in a future large scale experiment; show technical and engineering scalability toward a tonne-scale instrument; and perform searches for other physics beyond the Standard Model, such as dark matter and axions.

\begin {figure}[ht]
\includegraphics[width=0.45\textwidth]{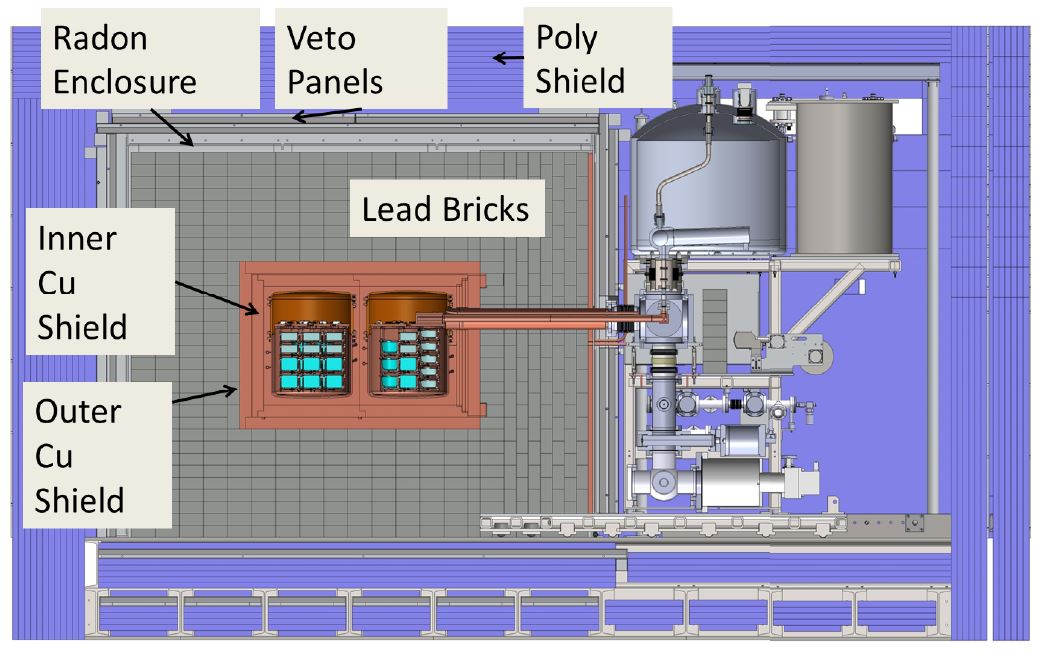}
\centering \caption{\it Schematic drawing of the \textsc{Majorana Demonstrator} shown with both modules installed.}
\label{fig:section}
\end {figure}

The experiment is composed of 44~kg of high-purity Ge detectors which act as source and detector of \ge\ \BBz. HPGe detectors benefit from the intrinsic low backgrounds of the source material, well-understood enrichment chemistry, and excellent energy resolution and event reconstruction capabilities. P-type point contact detectors~\cite{ppc,ppc2} were chosen after extensive R\&D by the collaboration for their powerful background rejection capabilities. The HPGe detectors operate under vacuum at liquid nitrogen temperature (77~K). Twenty nine kg of the detectors have been fabricated from Ge material that is enriched to $>$87\% in \ge\ and 15~kg from natural Ge (7.8\% \ge). The average mass of the enriched detectors is $\sim$850~g. A modular instrument composed of two cryostats, named Modules~1 and~2, built from ultra-pure electroformed copper is being constructed. Each module hosts 7 strings of 3-5 detectors. The strings are assembled and some of them are characterized in dedicated String Test Cryostats (STCs). The prototype module, an initial cryostat fabricated from commercially produced copper, took data with three strings of detectors produced from natural germanium. It served as a test bench for mechanical designs, fabrication methods, and assembly procedures to be used for the construction of the two electroformed-copper modules. The modules are operated in a passive shield that is surrounded by a 4$\pi$ active muon veto. To mitigate the effect of cosmic rays and prevent cosmogenic activation of detectors and materials, the experiment is being deployed at 4850~ft depth (4260~m.w.e. overburden) at the Sanford Underground Research Facility in Lead, SD~\cite{surf}. A schematic drawing of the \mj\ \dem\ is shown in Figure~\ref{fig:section}.

The background goals of the \dem~\cite{mjdbkg} require high-voltage (HV) cables for the~\mj~\dem~to be extremely low-mass miniature coaxial cables. They are produced by Axon'~Cable\footnote{http://www.axon-cable.com/} in cooperation with the~\mj~Collaboration. The outer diameter of the HV cables (Axon' part number TD11153B) is 1.2~mm, the conductor diameter is 0.152~mm and the inner dielectric has a diameter of 0.77~mm. The central conductor is a single conductor made of bare copper, the helical ground shield is made of 50AWG copper and the inner dielectric and outer jacket are made of extruded FEP (Fluorinated Ethylene Propylene). More infromation about the Axon' HV cables can be found in~\cite{MJDAssay}. The high voltage vacuum feedthroughs have to be small to integrate 80 connections in four flanges. At the same time, these cables and connectors must be capable of supplying HPGe detector operating voltages as high as 5~kV without exhibiting discharge that can damage the front-end electronics or mimic detector signals. The phenomenon of surface micro-discharge induced by HV and techniques for the reduction and discrimination of such breakdowns was discussed in~\cite{md,md2,mdATLAS} and references therein. Studies show that this discharge effect occurs mainly at interfaces, in microscopic voids between dielectric surfaces, on contaminated surfaces, and on surfaces with imperfections. The occurrence of these discharges in various materials and environments necessitated the unique HV component testing program described in this manuscript.

\begin {figure}[ht]
\subfigure[]{\includegraphics[width=0.33\textwidth]{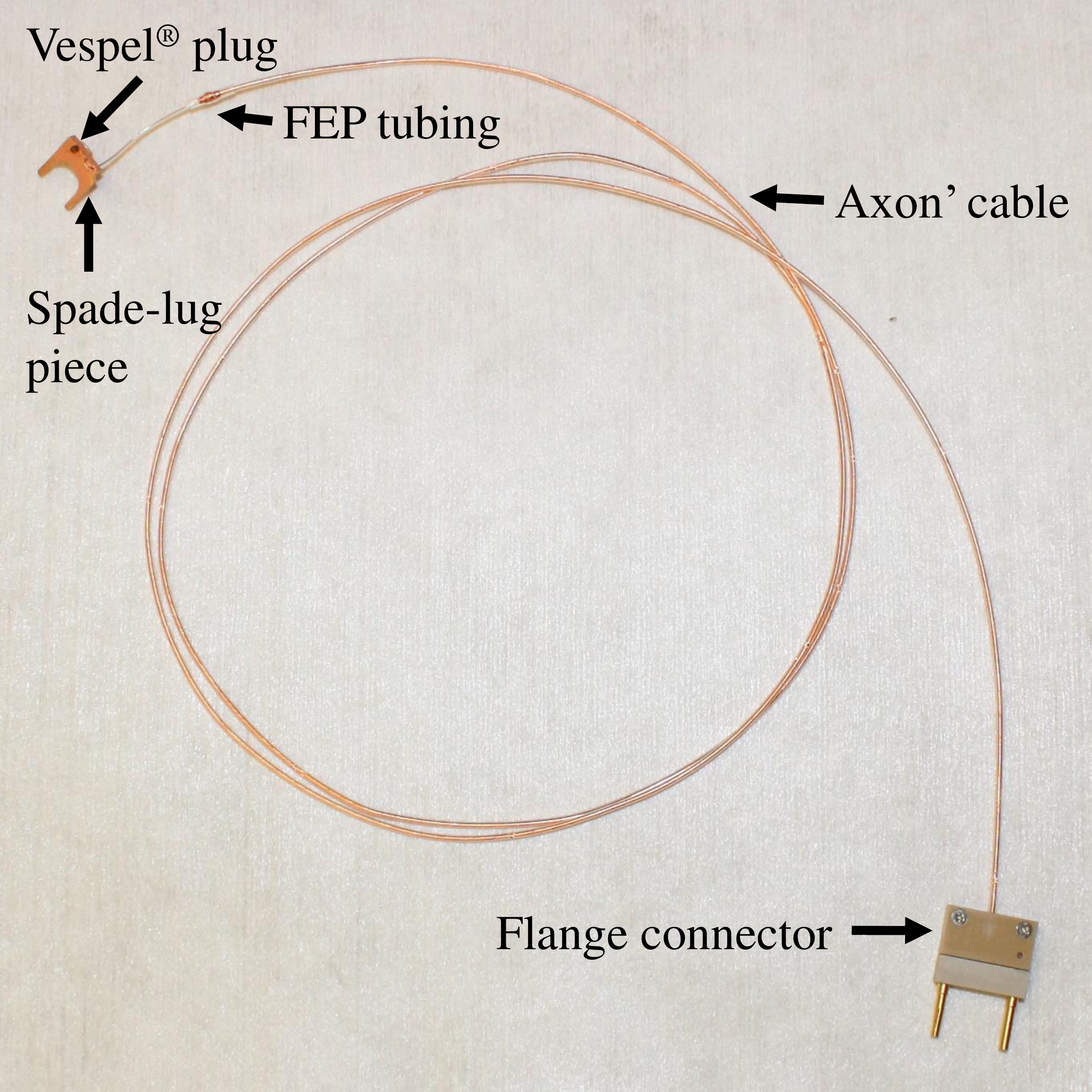}}
\subfigure[]{\includegraphics[width=0.33\textwidth]{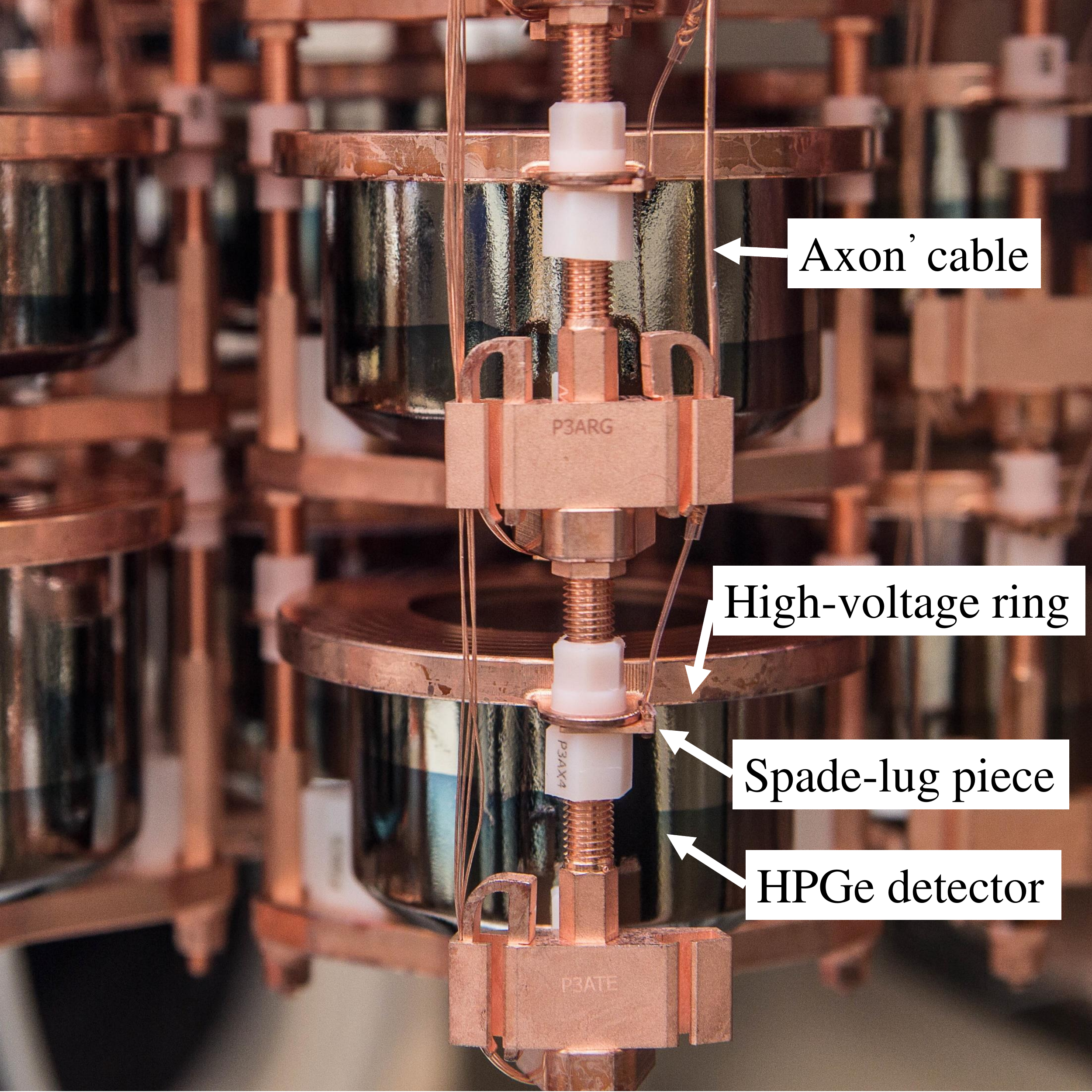}}
\centering \caption{\it (a) An Axon' HV cable with the connector that is attached to the flange pins (bottom right) and the copper spade-lug piece that makes contact with the copper HV ring at the detector (top left). The conductor is held in place by a Vespel$^{\circledR}$ plug and the ground is coiled and surrounded by FEP tubing. (b) Detectors in the~\mj~\dem~Module~1: The Axon' HV cables are on the right attached to the copper spade-lug piece in contact with the copper ring that provides high voltage to the detector.}
\label{fig:axon}
\end {figure}

A picture of one Axon' HV cable assembly is shown in Figure~\ref{fig:axon}(a) and a picture of detector strings in Module~1 of the~\mj~\dem\ with the cables and detectors installed is showed in Figure~\ref{fig:axon}(b). At the~\mj~\dem, the high voltage is provided to the detectors by external power supplies and brought into the vacuum flange mounted electrical feedthroughs by standard cables. Each Axon' HV cable is prepared from an 85~inch long strand. Both ends are first stripped to specifications. On the flange side, the outer dielectric is stripped by 0.5~inch while the inner dielectric is stripped by 0.25~inch. The 0.5~inch of exposed shielding is twisted together making sure no strands are left loose. The shielding and inner conductor are positioned in the sockets housed in a PEEK HV connector body where they are held in place with silver epoxy assuring electrical contact. The assembly is left to cure for 24~h before it is completed by closing the HV connector with a mating PEEK cover that provides strain relief. On the detector side, the outer dielectric is stripped by 1~inch while the inner dielectric is stripped by 0.13~inch. The 1~inch of exposed shielding is twisted together making sure no strands are left loose. It is then further twisted around the outer dielectric at the base of the cut and secured in place with a 0.25~inch long FEP heat shrink tube. The inner dielectric is threaded through a copper spade-lug piece that is in contact with a copper ring, which is on top of the detector and provides high-voltage. The cable is held in place at the spade-lug piece with a small Vespel$^{\circledR}$~SP-1\footnote{http://www.dupont.com/} plug that prevents the conductor cable from being exposed, preventing discharge.


The setup devoted to testing of the cables and feedthroughs is described in Section~\ref{sec2}. Section~\ref{sec3} reports the R\&D process that led to the optimum configuration, ensuring the correct performance of the HV cables and feedthrough flanges. All of the HV cables and feedthrough flanges to be used at the~\mj~\dem~were characterized in this setup prior to installation; the results are presented in Section~\ref{sec4}. The micro-discharge effects in the~\mj~\dem~data are estimated in Section~\ref{sec5} and a projection for a large scale experiment is presented in Section~\ref{sec6}.

\section{Experimental setup}
\label{sec2}

\begin {figure}[ht]
\includegraphics[width=0.45\textwidth]{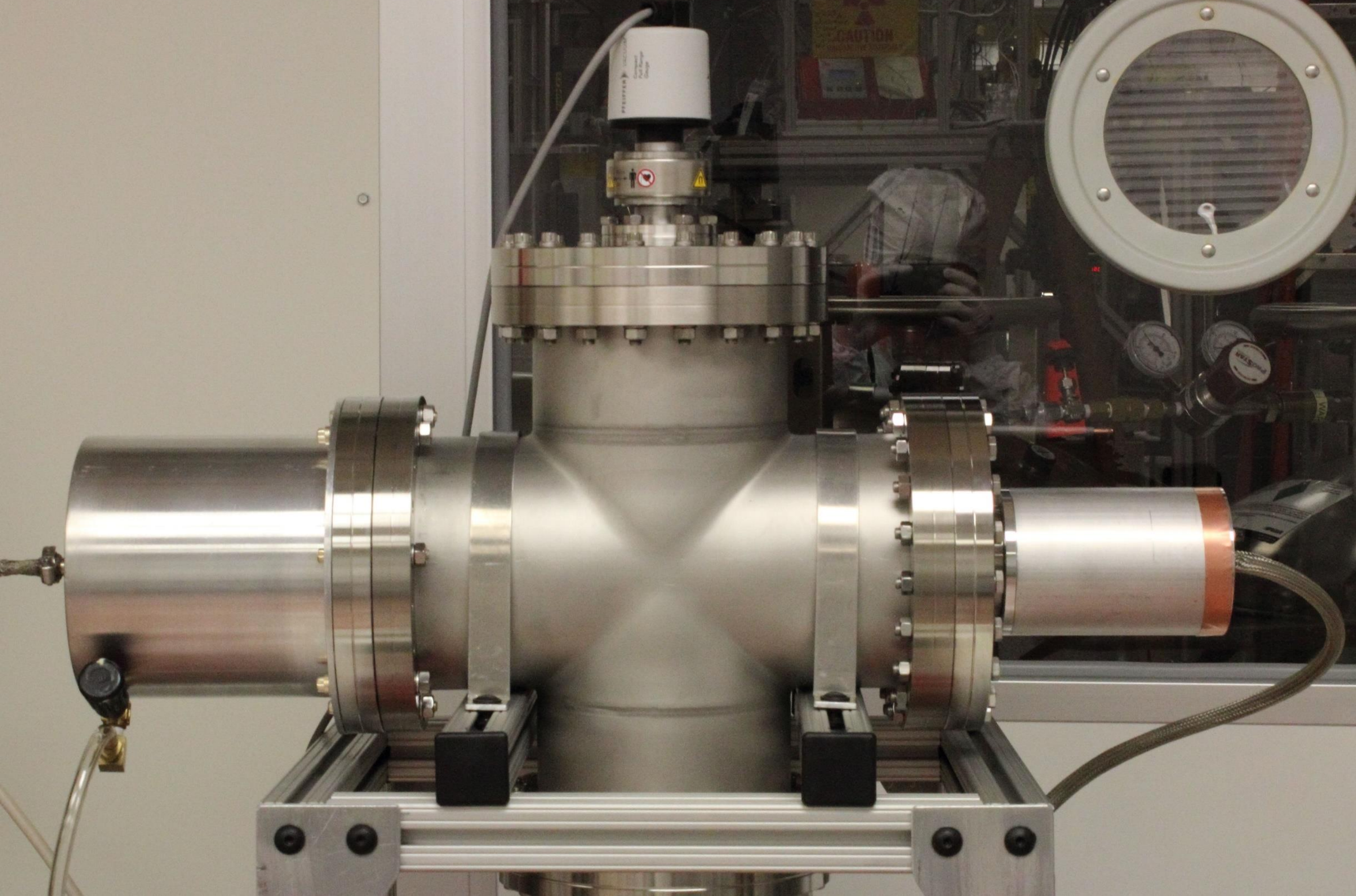}
\centering \caption{\it Testing chamber used in the HV measurements described in this work. The flange to be tested is mounted on the left port. It is shielded by a metal case which also allows for N$_{2}$ to be blown onto the flange. On the right, a 30 kV feedthrough is mounted on a 2.75~inch flange to be used to test cables independently of a feedthrough flange. The pressure gauge is mounted on top, and the vacuum pump is attached at the bottom (not shown).}
\label{fig:chamber}
\end {figure}

\begin {figure}[ht]
\includegraphics[width=0.48\textwidth]{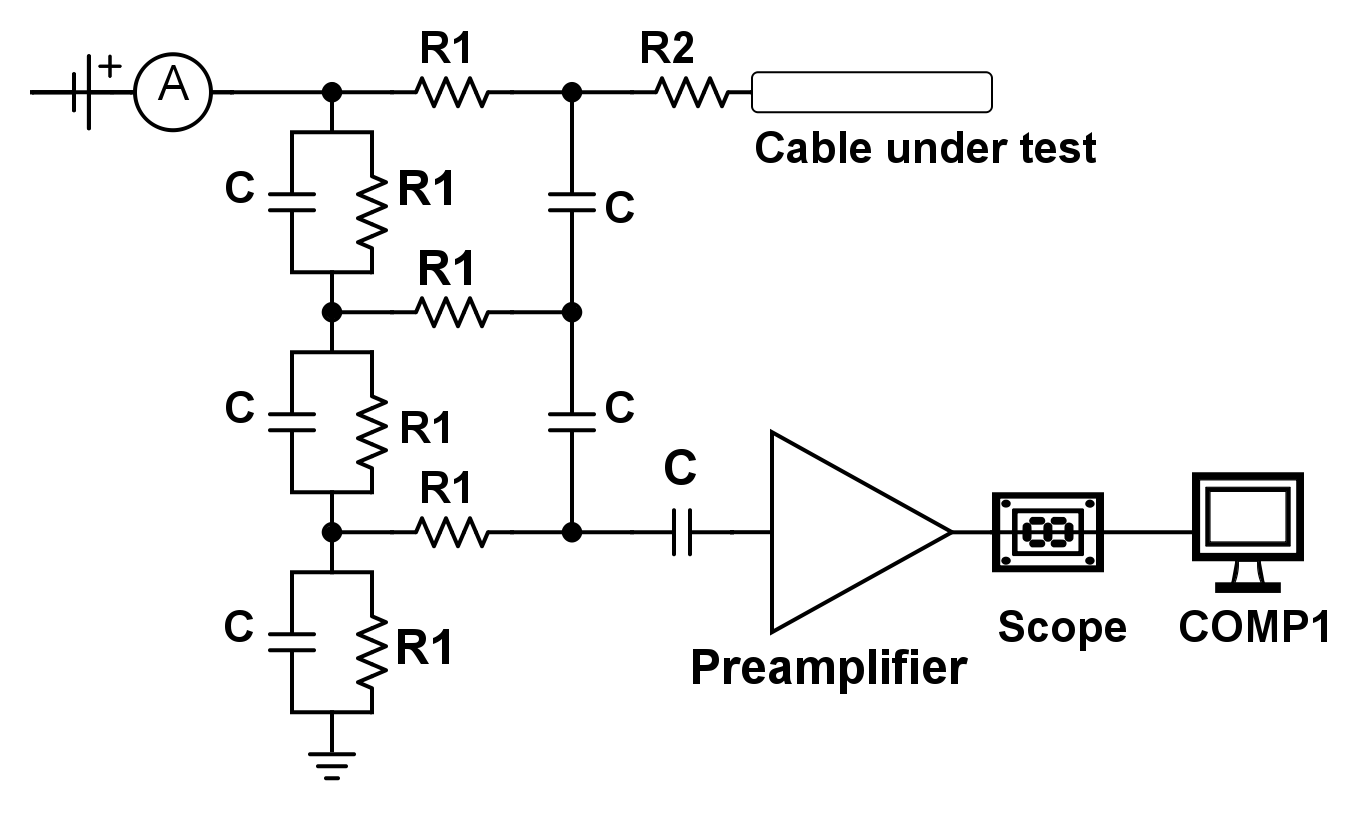}
\centering \caption{\it Schematic drawing of the experimental setup. High voltage is provided by a power supply and the output is fed to a high voltage-conditioning stack composed by a set of resistances (R1\,=\,1~G$\Omega$ and R2\,=\,503~$\Omega$), capacitors (C\,=\,0.01~$\mu$F) and corona disks (drawn as black dots). The cable being tested gets charged through the stack and if a discharge occurs on it, the output is fed into the preamplifier. The output of the preamplifier is fed to an oscilloscope and triggers are recorded by a PC running Labview software.}
\label{fig:diagram}
\end {figure}

A dedicated setup was designed and constructed at the University of Washington to accommodate testing of the cables and feedthroughs under vacuum. The setup reproduces as closely as possible the full HV cable path as designed for the~\mj~\dem. The cables are tested in an all-metal vacuum chamber shown in Figure~\ref{fig:chamber}, evacuated by a Pfeiffer HiPace 80 turbomolecular pump\footnote{http://www.pfeiffer-vacuum.com/}. The base pressure is approximately 10$^{-7}$~mbar. A schematic drawing of the experimental setup is shown in Figure~\ref{fig:diagram}. Positive high voltage is provided by a Bertran 375P power supply with the current independently read out by a Hewlett Packard 3458A multimeter. The output of the power supply is fed to a high voltage-conditioning stack composed by a set of resistances, capacitors and corona disks to avoid discharges. The cable under testing gets charged through the stack and if a discharge occurs on it, the output is fed into the preamplifier. The preamplifier was used in the SNO neutral-current detectors and is described in~\cite{SNO}. Both the HV stack and the preamplifier are housed into a radio-frequency electromagnetic radiation (RF) shielded box via standard HV cable. The output of the preamplifier is connected to a Tektronix TDS2024B oscilloscope read out by a PC running LabVIEW software. The oscilloscope trigger level was -15~mV, note that the micro-discharges constitute a voltage drop measured as negative going pulses. The initial R\&D measurements explained in Section~\ref{sec3} were carried out in standard laboratory conditions. Before taking the characterization measurements explained in Section~\ref{sec4}, the system was moved to a clean room to operate in the adequate cleanliness conditions required for handling the~\mj~\dem\ components.

\begin {figure}[ht!]
\subfigure[]{\includegraphics[width=0.463\textwidth]{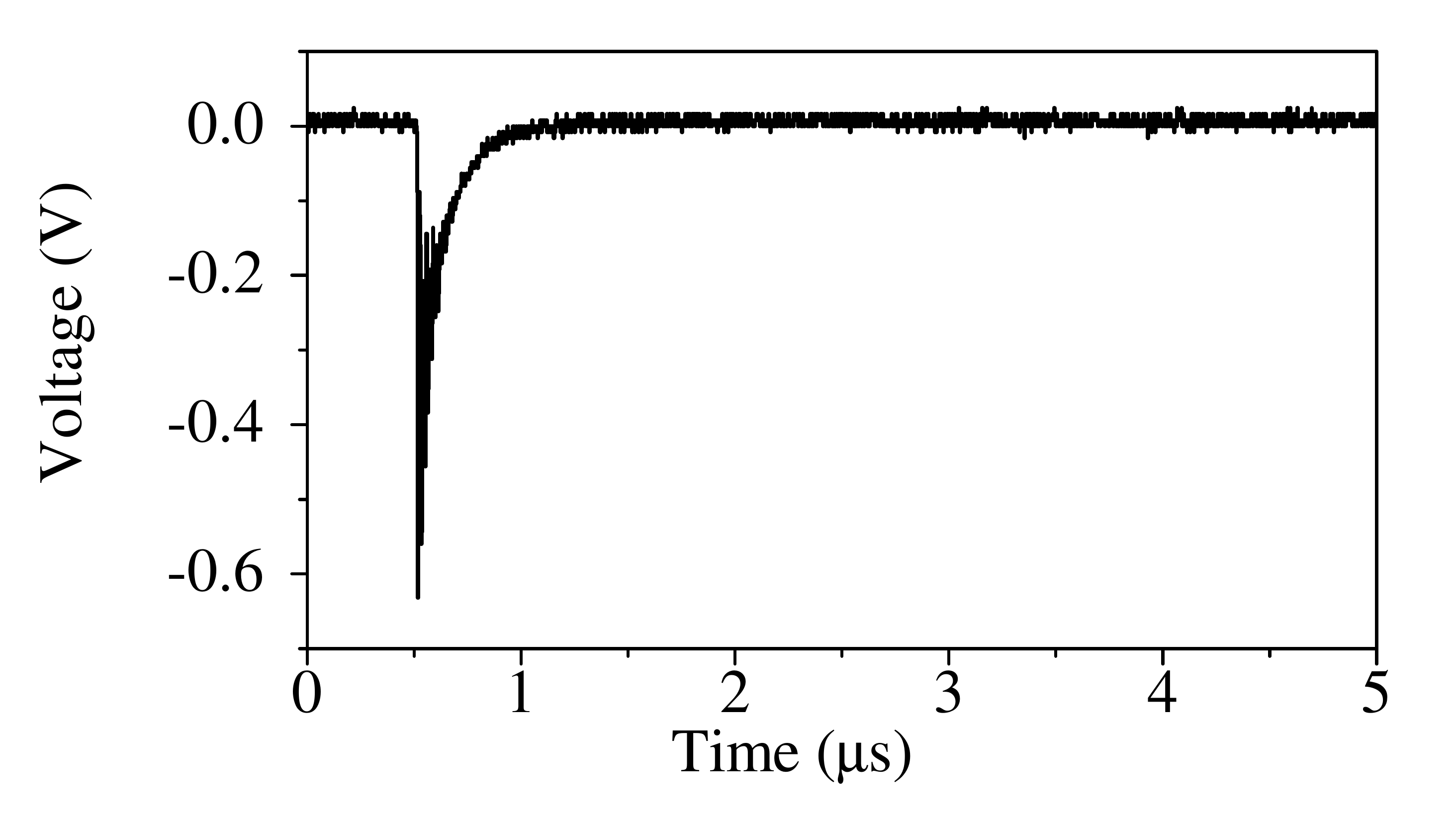}}
\subfigure[]{\includegraphics[width=0.463\textwidth]{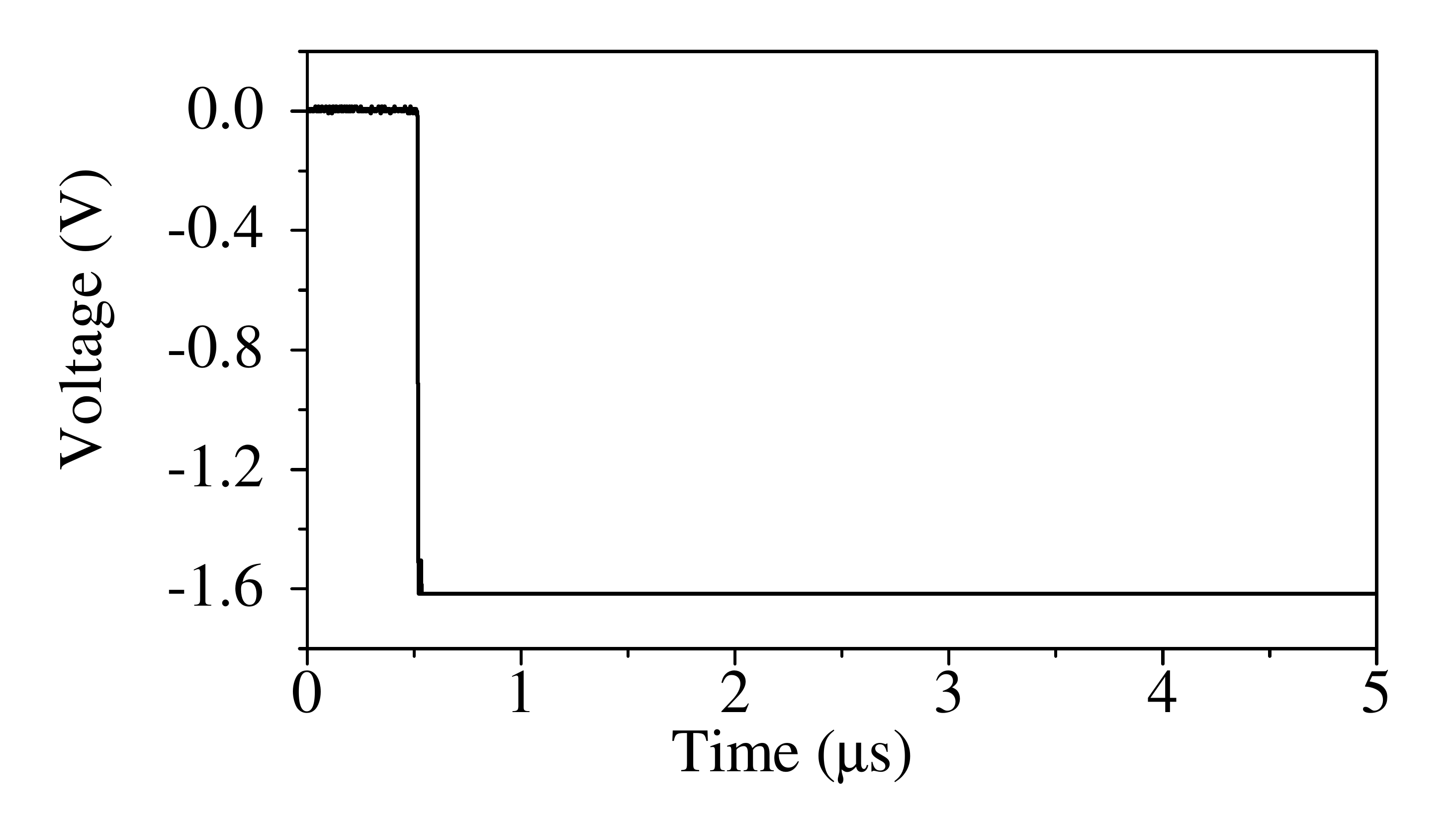}}
\subfigure[]{\includegraphics[width=0.463\textwidth]{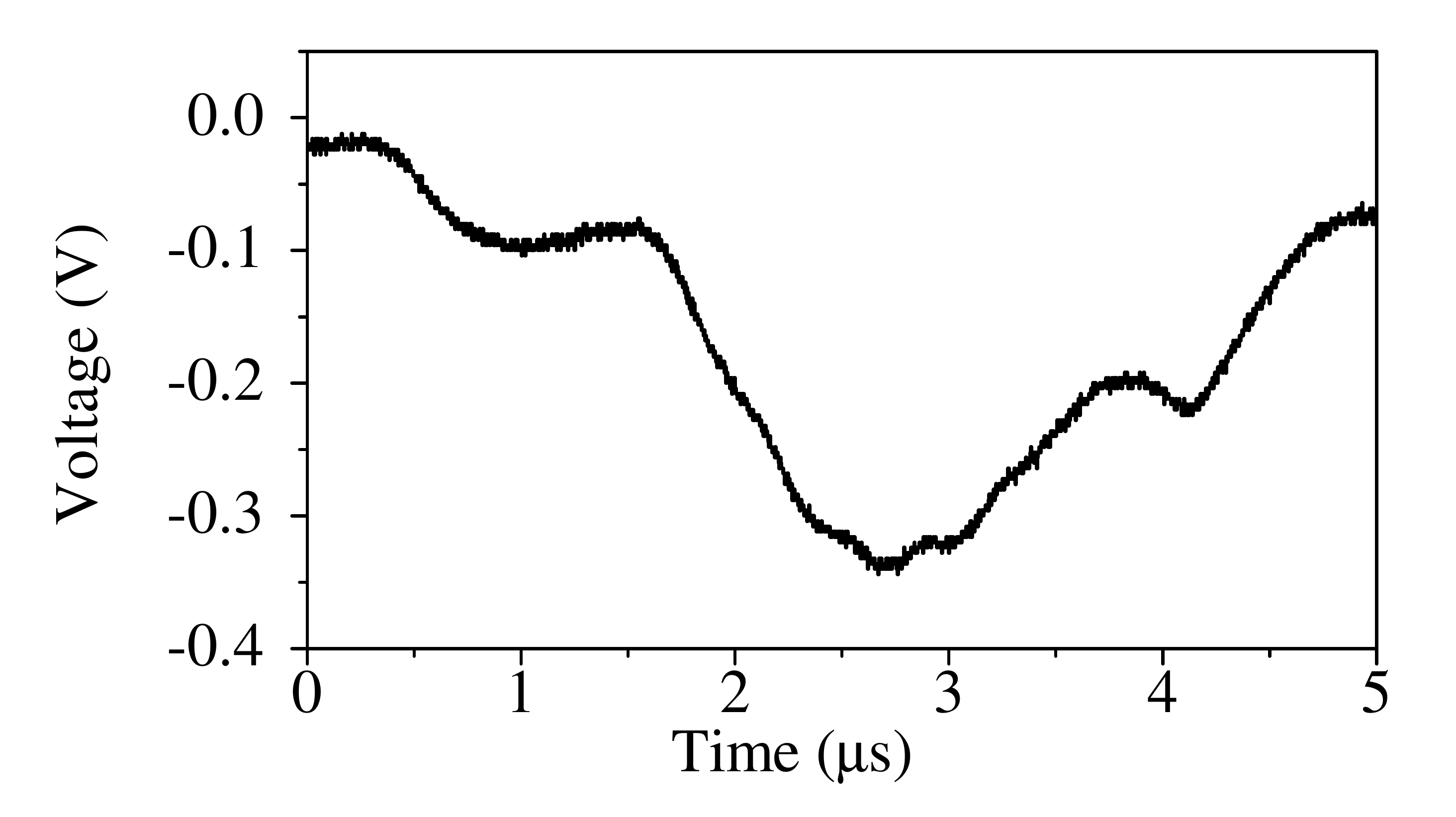}}
\subfigure[]{\includegraphics[width=0.463\textwidth]{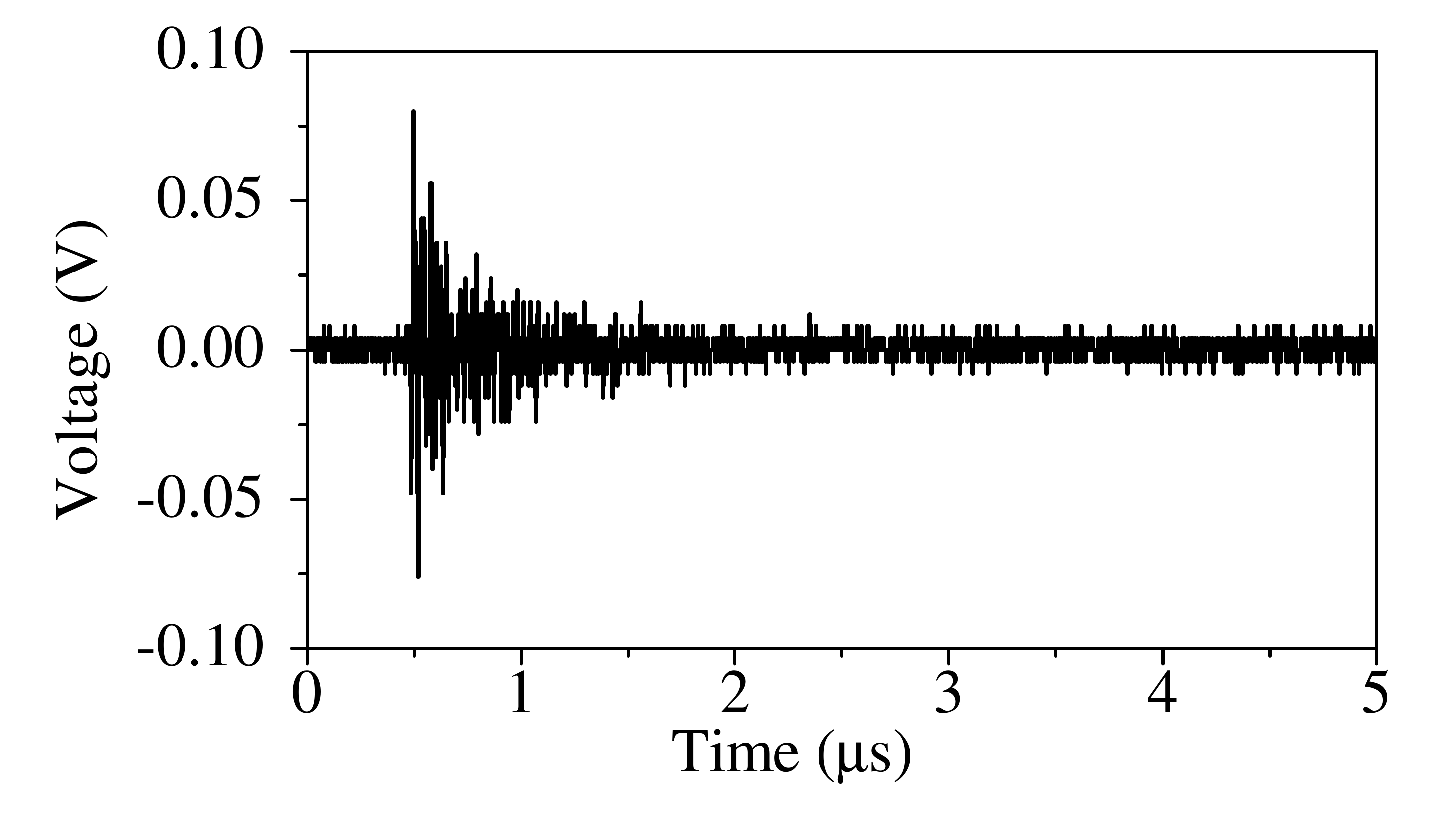}}
\centering \caption{\it Examples of the different waveforms categorized: micro-discharge (a), big micro-discharge (b), breakdown (c) and transient noise (d).}
\label{fig:pulses}
\end {figure}

The recorded pulse shapes were divided into four categories. Thanks to the low rate of events, the events are individually inspected and categorized by the analyzers. Examples of these events are shown in Figure~\ref{fig:pulses}:
 \begin{enumerate}
   \item Micro-discharge events ($\mu$d): Fast negative events created by a small micro-discharge in the testing setup.
   \item Big micro-discharge events (b$\mu$d): Micro-discharges where the pulse saturated both scales of the scope, which were set at ~-1.6~V and 5~$\mu$s. They are potentially more dangerous than micro-discharges because they are bigger and could damage the cable and/or the electronics although their origin could be the same.
   \item Breakdown events: Those events exhibiting strong negative baseline fluctuations induced by charge, often with slower micro-discharge events superimposed. They are dangerous since a long exposure to this kind of event can damage the cable and/or the electronics.
   \item Noise events: Transient pickup noise due to imperfect shielding of the RF shielded box and the cables and connectors in the setup. It is identified by its symmetry about the X axis. These events are not relevant for this work and were discarded in the analysis.
 \end{enumerate}

\begin {figure}[ht]
\includegraphics[width=0.25\textwidth]{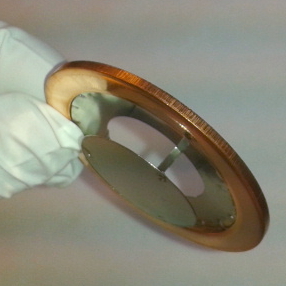}
\centering \caption{\it Picture of the baffle used to avoid the micro-discharges created by the gauge.}
\label{fig:baffle}
\end {figure}

It is worth mentioning that hints of a systematic source of micro-discharge events in the testing system were found initially. A higher rate of micro-discharge events was observed when the cable was under vacuum than when it was tested in air. In addition, a PEEK connector without a cable plugged into the vacuum feedthroughs was able to create micro-discharges. This led to the hypothesis that these micro-discharges were created by the cold-cathode ion gauge used to measure pressures in the vacuum system~\cite{vacuum}. The gauge can be seen in Figure~\ref{fig:chamber} on top of the vacuum chamber. To test the gauge, an Axon' cable terminated by a smooth brass sphere was laid over a nylon block inside the vacuum chamber. With the gauge on, a rate of 21\,$\pm$\,1\,$\mu$d/h was detected; whereas a rate of 0\,$\mu$d/h was reported with the gauge off. The conclusion was that ions emitted by the gauge are able to charge the insulator; when the static charge buildup discharges, micro-discharge events are recorded. This effect was avoided by placing a baffle, shown in Figure~\ref{fig:baffle}, near the gauge. It prevents the emitted ions from reaching the parts being tested. The baffle was tested with the same cable and the gauge on, giving a result of 0\,$\mu$d/h. This effect was discovered after the R\&D measurements presented in Section~\ref{sec3}, and rates presented there can have a contribution from it, but the baffle was subsequently used in the characterization measurements shown in Section~\ref{sec4}.

\section{R\&D measurements}
\label{sec3}

At first, measurements were carried out to understand the testing system and determine the optimum design configuration. As a result of these R\&D measurements, improvements to the cable layout were made and the best feedthrough flange model was selected.

\subsection{Cable layout}
\label{sec3.1}
 Different cable layouts were tested individually and in various configurations. In these tests the HV was brought into the chamber by a 30 kV feedthrough mounted on a 2.75~inch ConFlat flange. The feedthrough was terminated on the vacuum side by a smooth brass sphere, into which the stripped end of the cable under testing was inserted. These were the different tests performed and conclusions drawn:
\begin{itemize}
 \setlength\itemsep{0em}
  \item The voltage dependence was studied with an Axon' cable. Results are shown in Table~\ref{tab:voltage}. The individual errors showed there, and in the following, are calculated as the square root of the total number of micro-discharges divided by the sampling time. It was observed that reducing the voltage also reduces the micro-discharge rate. 
  \item The coiling of the cable was discovered to be an important factor to take into account. A rate of 17\,$\pm$\,0.6\,$\mu$d/h was measured with an Axon' cable wrapped in a tangled and tight way, whereas a rate of 0.03\,$\pm$\,0.03\,$\mu$d/h had been measured before with the same cable coiled with a radius more than 10 times greater than the cable radius. As a conclusion, the bending radius is recommended to be at least 10 times greater than the cable radius to avoid breakdown-inducing structural damage.
  \item The strip-back length (separation distance between shield and conductor) was also studied, see Table~\ref{tab:sbl}. We observed that breakdown occurred when the strip-back length was less than 0.5~inches. It was decided to add FEP tubing surrounding the ground, and to maintain a strip-back length of 1~inch.
  \item The copper spade-lug piece was tested with a standard RG59 HV cable. It was observed that when the piece was in contact with a nylon block used as a support the micro-discharge rate was 4.7\,$\pm$\,0.3\,$\mu$d/h. However no micro-discharge events were reported when it was hanging under vacuum. Electropolishing the piece successfully reduced the micro-discharge rate to 0.05\,$\pm$\,0.03\,$\mu$d/h, but was determined to not be required in the actual configuration of the~\mj~\dem.
\end{itemize}

\begin{table}[ht]
\begin{center}
\fboxrule=0cm \fboxsep=0cm \fbox{
\begin{tabular}[]{ll}
\toprule
Voltage & $\mu$d rate  \\
\cmidrule{1-2}															
 5.0 kV	& 10.0\,$\pm$\,0.7\,$\mu$d/h    \\
 4.9 kV	& 3.85\,$\pm$\,0.25\,$\mu$d/h    \\
 4.8 kV	& 0.43\,$\pm$\,0.12\,$\mu$d/h    \\
 4.7 kV	& 0.15\,$\pm$\,0.04\,$\mu$d/h    \\
\bottomrule
\end{tabular}
}\caption{\it Voltage dependence of micro-discharge rate measured for a test Axon' cable.}
\label{tab:voltage}
\end{center}
\end{table}

\begin{table}[ht]
\begin{center}
\fboxrule=0cm \fboxsep=0cm \fbox{
\begin{tabular}[]{lll}
\toprule
Strip-back length & FEP tubing & $\mu$d rate  \\
\cmidrule{1-3}															
 0.50 inch	& No & Breakdown    \\
 1.00 inch	& No & 1390\,$\pm$\,60\,$\mu$d/h    \\
 1.25 inch	& No & 1.58\,$\pm$\,0.29\,$\mu$d/h    \\
 1.50 inch	& No & 0.12\,$\pm$\,0.08\,$\mu$d/h    \\
 0.30 inch	& Yes &  Breakdown    \\
 1.00 inch	& Yes & 0.23\,$\pm$\,0.10\,$\mu$d/h    \\
\bottomrule
\end{tabular}
}\caption{\it Results of the different strip-back length values tested.}
\label{tab:sbl}
\end{center}
\end{table}

\subsection{Flange selection}
\label{sec3.2}

\begin {figure}[ht]
\subfigure[]{\includegraphics[width=0.22\textwidth]{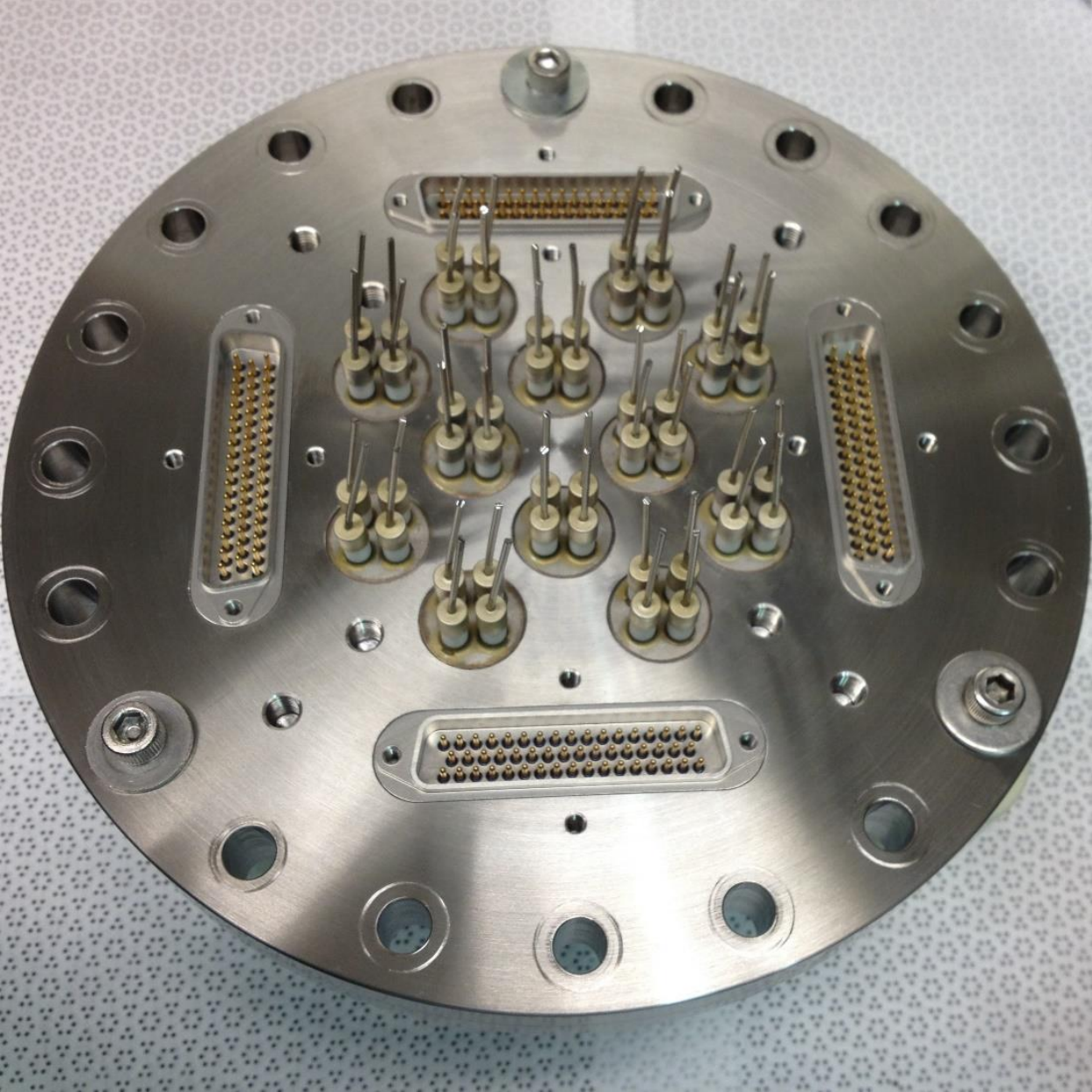}}
\subfigure[]{\includegraphics[width=0.22\textwidth]{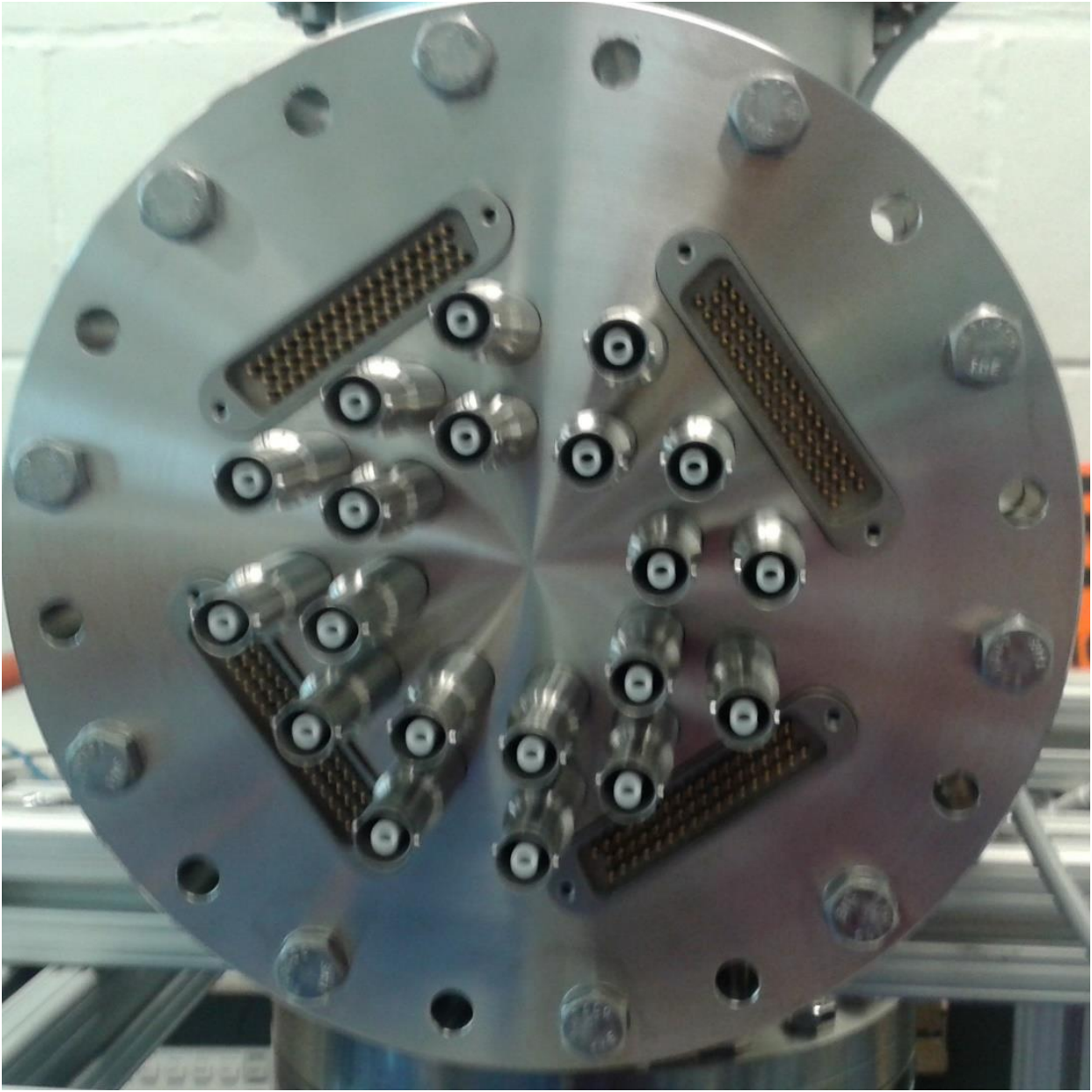}}
\subfigure[]{\includegraphics[width=0.22\textwidth]{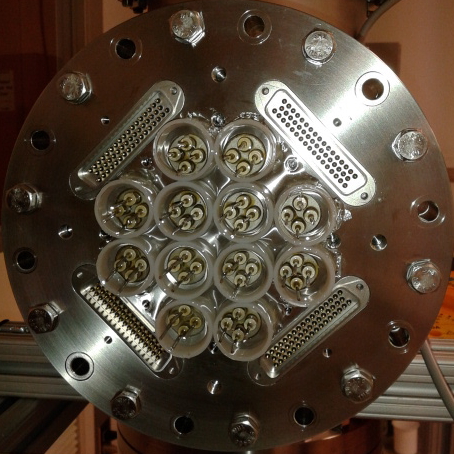}}
\subfigure[]{\includegraphics[width=0.22\textwidth]{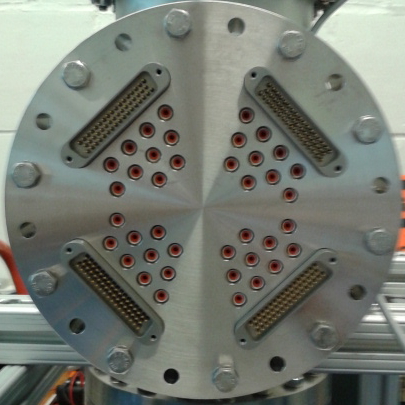}}
\centering \caption{\it Pictures of the different flange designs considered to be used at the~\mj~\dem. (a) Ceramic feedthrough design, (b) SHV connectors, (c) ceramic feedthrough design potted and (d) Pee-Wee connectors from SRI Hermetics.}
\label{fig:flanges}
\end {figure}

The first flange design, shown in Figure~\ref{fig:flanges}(a) was based on ceramic feedthroughs clustered in groups of four. When tested, this design exhibited current fluctuations at 5\,kV when HV and ground were applied to neighboring pins. The source of this failure was tracked to the manufacturer designing for a pin-to-pin breakdown voltage of 1\,kV when 5\,kV was specified. It was only possible to operate it under a N$_{2}$ purge to reduce humidity.

Three different replacement flange designs were studied in parallel: a flange with SHV connectors (see Figure~\ref{fig:flanges}(b)), potting the original ceramic feedthrough flange design (see Figure~\ref{fig:flanges}(c)), and a flange with a new type of feedthrough, called ``Pee-Wee" connectors (see Figure~\ref{fig:flanges}(d)) from SRI Hermetics\footnote{http://www.srihermetics.com/}. Two flanges of each model were tested, 1~x~8CF (for the \mj\ \dem) and 1~x~6CF diameter (for the STCs). Only one feedthrough was tested at a time leaving the others disconnected and without anything connected to the vacuum side. Average leakage current measurement and standard deviation values are shown in Table~\ref{tab:dc}. Only two feedthroughs with leakage current $>$2~$\mu$A were found in any of the three designs, one feedthrough in a potted flange and another in a Pee-Wee flange; both affected feedthroughs could still be used as ground since in these designs one half of the pins are used as HV and the other half as ground, which are not sensitive to micro-discharges. Some of the feedthroughs of each flange were tested for micro-discharge. Small micro-discharge events were reported on all three flanges (see rates in Table~\ref{tab:flanges}). The uncertainty in the average rate is calculated by propagating the individual uncertainties. They are considered to be uncorrelated, and the same procedure is used in the following when showing average rate values. A few big micro-discharge events were observed with the Pee-Wee flange but not in the other two. No breakdown was found in any of the flanges. Ultimately, all three flange models were certified to be used. The SHV and potted flanges were rejected for technical reasons: the long stand-off length of the SHV flange connectors was difficult to interface to the  already built external preamplification electronics hardware, and the long-term behavior of the potted material was not guaranteed in the planned deployment time of the experiment. The potted flanges were used in the prototype module since they were needed earlier and its planned lifetime was short. Their performance was good, more details can be found in Section~\ref{sec5}. Finally, the Pee-Wee feedthrough flange model was selected for use in the~\mj~\dem. 

\begin{table}[ht]
\begin{center}
\fboxrule=0cm \fboxsep=0cm \fbox{
\begin{tabular}[]{llll}
\toprule
Flange model & Pins & High I$_{\rm leak}$ & Average I$_{\rm leak}$   \\
\cmidrule{1-4}															
 SHV	& 25&0 & 1547\,$\pm$\,2\,nA      \\
 Potted	& 60&1 & 1559\,$\pm$\,28\,nA     \\
 Pee-Wee	& 50&1 & 1555\,$\pm$\,14\,nA     \\
\bottomrule
\end{tabular}
}\caption{\it Leakage current (I$_{\rm leak}$) measurement results for the three flange models considered for the \textsc{Majorana Demonstrator}. Pins refers to the number of pins tested. High I$_{\rm leak}$ refers to the number of tested pins exhibiting a too high leakage current to be used for HV connections. The average leakage current and corresponding standard deviation of the tested pins are presented.}
\label{tab:dc}
\end{center}
\end{table}

\begin{table}[ht]
\begin{center}
\fboxrule=0cm \fboxsep=0cm \fbox{
\begin{tabular}[]{lll}
\toprule
Flange model & Pins & Average $\mu$d rate   \\
\cmidrule{1-3}															
 SHV	& 2& 3.67\,$\pm$\,0.29\,$\mu$d/h      \\
 Potted	& 3& 7.23\,$\pm$\,0.31\,$\mu$d/h     \\
 Pee-Wee	& 4& 0.58\,$\pm$\,0.07\,$\mu$d/h     \\
\bottomrule
\end{tabular}
}\caption{\it Average micro-discharge rate measured for the three flange models considered to be used in the \textsc{Majorana Demonstrator} during the cursory measurements performed to aid in selecting a feedthrough design. The number of pins tested is shown in the second column.}
\label{tab:flanges}
\end{center}
\end{table}




\section{Characterization measurements}
\label{sec4}

\begin {figure}[ht]
\includegraphics[width=0.45\textwidth]{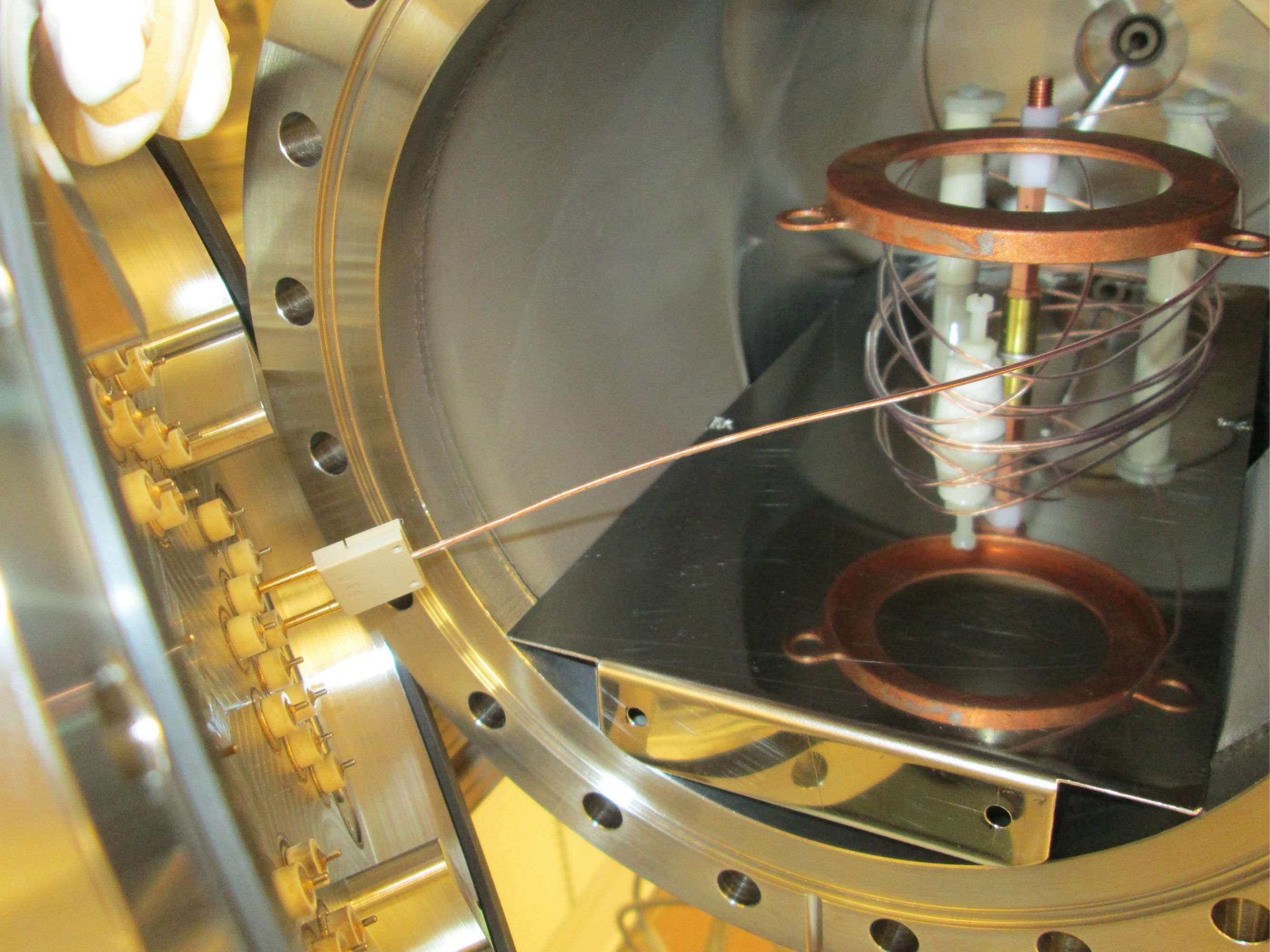}
\centering \caption{\it Picture of an HV cable and flange being characterized inside the testing chamber.}
\label{fig:chamber2}
\end {figure}

All cables and feedthroughs, including spares, to be used in the~\mj~\dem\ modules and STCs were tested after moving the HV testing apparatus to a clean room at the University of Washington. In total, 133 HV cables were characterized (42 for Module~1, 42 for Module~2, and 49 for the STCs), as were 280 HV feedthroughs (five 8-inch flanges with 40 feedthroughs each and eight 6-inch flanges with 10 feedthroughs each). A picture of a cable and a flange being tested can be seen in Figure~\ref{fig:chamber2}. The cable was tested with the copper spade-lug piece attached to it and connected to the HV ring. The cable attached to the HV ring was placed in a test stand inside the testing chamber and connected to the flange.

First, the leakage current of the feedthroughs was measured up to 5~kV. Feedthroughs exhibiting current greater than 2~$\mu$A were marked for use as ground in the~\mj~\dem\ and were not tested further. The results are shown in Table~\ref{tab:dctesting}, where it can be seen that 97\% of the pins showed leakage current values $<$2~$\mu$A at 5~kV. The average values are higher than those shown in Table~\ref{tab:dc} because the external HV cable was different in these measurements. In parallel, the cables were tested on a low-sensitivity HV test bench to check for breakdowns. Each cable was biased up to 5.2~kV while the trip current of the HV power supply is set to 80~nA on the most sensitive scale. All the cables passed the test, i.e. no current trips were found.

\begin{table}[ht]
\begin{center}
\fboxrule=0cm \fboxsep=0cm \fbox{
\begin{tabular}[]{llll}
\toprule
Flange model & Pins & High I$_{\rm leak}$ & Average I$_{\rm leak}$   \\
\cmidrule{1-4}															
 6 inch & 80&5  & 1643\,$\pm$\,63\,nA       \\
 8 inch	& 200&2 & 1653\,$\pm$\,57\,nA     \\
\bottomrule
\end{tabular}
}\caption{\it Results of leakage current (I$_{\rm leak}$) measurements of the flange pins of the \mj\ \dem. The average leakage current and corresponding standard deviation of the tested pins are presented. Pins refers to the number of pins tested .}
\label{tab:dctesting}
\end{center}
\end{table}

\begin {figure}[ht]
\includegraphics[width=0.45\textwidth]{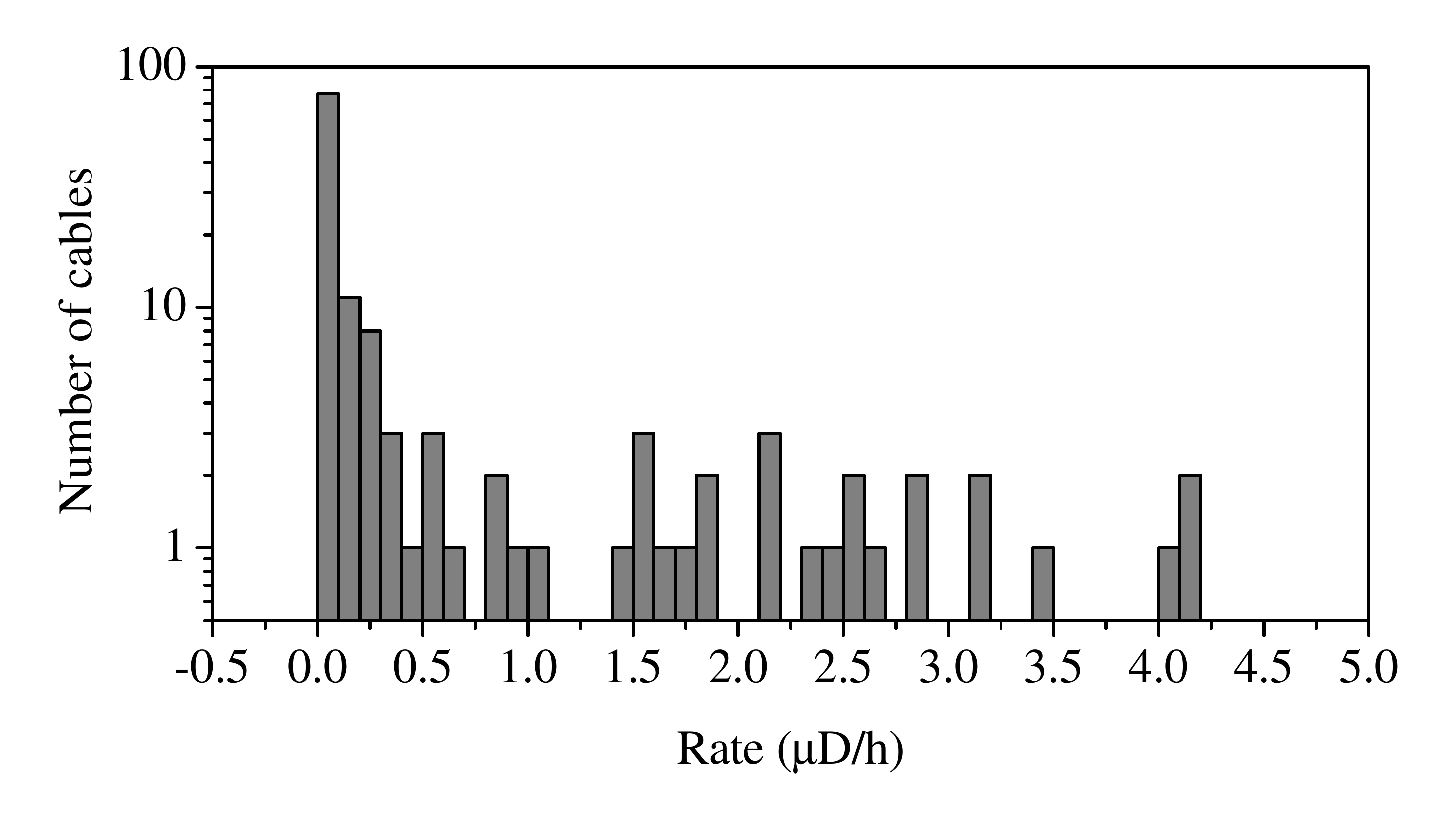}
\centering \caption{\it Micro-discharge rate measured for the cables to be used in the \textsc{Majorana Demonstrator}.}
\label{fig:cables}
\end {figure}

\begin {figure}[ht]
\includegraphics[width=0.45\textwidth]{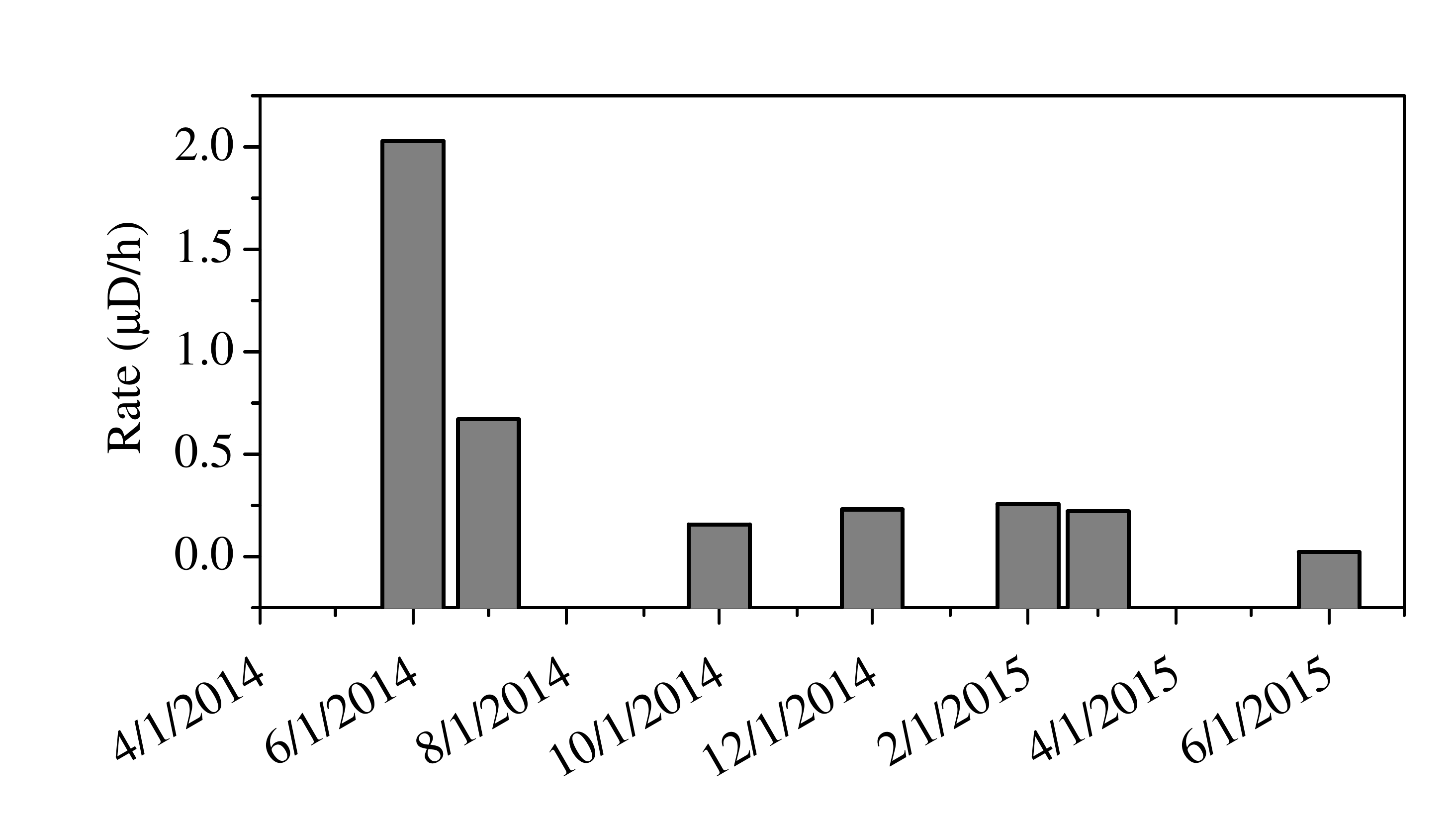}
\centering \caption{\it Micro-discharge rate per cable averaged per month.}
\label{fig:time}
\end {figure}

Second, each cable was connected to a feedthrough and characterized at 5~kV for an average of 14~hours (and a minimum time of 3~hours). In order to be accepted for use a cable must have a micro-discharge rate below 10~$\mu$d/h and big micro-discharge rate below than 0.1~b$\mu$d/h. This limit was set to avoid big rates that could introduce dead time to the experiment or damage the front-end electronics. If they failed the test, the cable and the feedthrough were tested separately to identify which was causing the high micro-discharge rate. Results of the micro-discharge rate per cable are shown in Figure~\ref{fig:cables} and Table~\ref{tab:cables}. The rate should be considered an upper limit since a contribution to the rate from the feedthrough or other parts of the electronic chain is expected. No issues were found with any of the cables, and a rate less than 5~$\mu$d/h was measured for all cables (with $\sim$5\% contribution from B$\mu$d's). The measurements were taken in the order displayed in Table~\ref{tab:cables}. The average micro-discharge rate is lower in the last measurements pointing to some micro-discharge events coming from other parts of the system and related to a long-term conditioning rather than only from the cables or feedthroughs. This effect can be observed in Figure~\ref{fig:time} where the measurements taken every month have been averaged.

\begin{table}[ht]
\begin{center}
\fboxrule=0cm \fboxsep=0cm \fbox{
\begin{tabular}[]{lll}
\toprule
Use & Cables & Average $\mu$ rate   \\
\cmidrule{1-3}															
 STC	& 42& 0.96\,$\pm$\,0.07\,$\mu$d/h      \\						
    	&  & 0.054\,$\pm$\,0.008\,b$\mu$d/h      \\
 Module 1	& 49& 0.67\,$\pm$\,0.04\,$\mu$d/h     \\
     	&  & 0.035\,$\pm$\,0.009\,b$\mu$d/h      \\
 Module 2	& 49 & 0.22\,$\pm$\,0.02\,$\mu$d/h     \\
    	&  & 0.012\,$\pm$\,0.004\,b$\mu$d/h      \\
 \cmidrule{1-3}
 Total	& 133& 0.62\,$\pm$\,0.03\,$\mu$d/h     \\
     	&  & 0.034\,$\pm$\,0.004\,b$\mu$d/h      \\
\bottomrule
\end{tabular}
}\caption{\it Average micro-discharge rate per cable measured for the cables to be used at the \textsc{Majorana Demonstrator}. }
\label{tab:cables}
\end{center}
\end{table}

\begin {figure}[ht]
\includegraphics[width=0.45\textwidth]{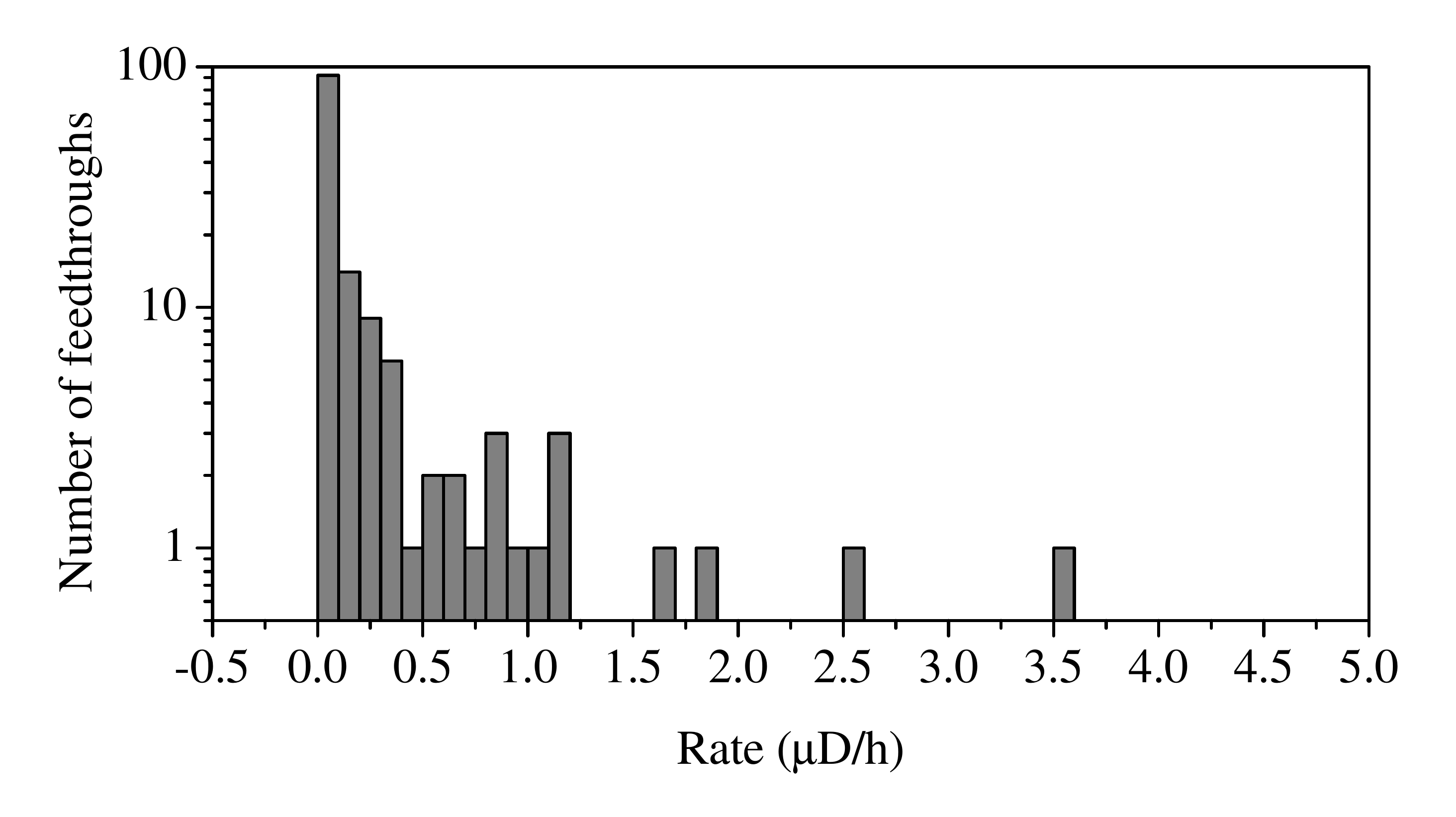}
\centering \caption{\it Micro-discharge rate measured for the feedthroughs to be used at the \textsc{Majorana Demonstrator}.}
\label{fig:flanges2}
\end {figure}

The results displayed per feedthrough can be seen in Figure~\ref{fig:flanges2} and Table~\ref{tab:flanges2}. In this case, the rate should also be considered an upper limit since a contribution to the rate from the cable or other parts of the electronics chain is expected. The rate at one feedthrough was 10.3\,$\pm$\,1.3\,$\mu$d/h, so it was recommended that that channel be used for a ground connection.

\begin{table}[ht]
\begin{center}
\fboxrule=0cm \fboxsep=0cm \fbox{
\begin{tabular}[]{lll}
\toprule
Flange model & Pins & Average $\mu$d rate   \\
\cmidrule{1-3}															
6 inch 	& 35& 0.17\,$\pm$\,0.02\,$\mu$d/h      \\
    	&  & 0.015\,$\pm$\,0.005\,b$\mu$d/h      \\
8 inch	& 100& 0.23\,$\pm$\,0.05\,$\mu$d/h     \\
    	&  & 0.08\,$\pm$\,0.02\,b$\mu$d/h      \\
 \cmidrule{1-3}
 Total	& 135& 0.20\,$\pm$\,0.03\,$\mu$d/h     \\
    	&  & 0.048\,$\pm$\,0.010\,b$\mu$d/h      \\
\bottomrule
\end{tabular}
}\caption{\it Average micro-discharge rate per pin measured for the feedthroughs to be used at the \textsc{Majorana Demonstrator}. }
\label{tab:flanges2}
\end{center}
\end{table}

\section{Micro-discharge effect in the \mj\ \dem}
\label{sec5}

A configuration with a low expected micro-discharge rate and no breakdowns has been found and no issues are expected in the \mj\ \dem. In addition, the micro-discharge effect in the \mj\ \dem\ will not affect the physics measurements because the polarity of waveforms induced at the charge collection electrode of the germanium detectors as a result of micro-discharges is opposite to that of waveforms associated with physics events.

\begin {figure}[ht]
\includegraphics[width=0.45\textwidth]{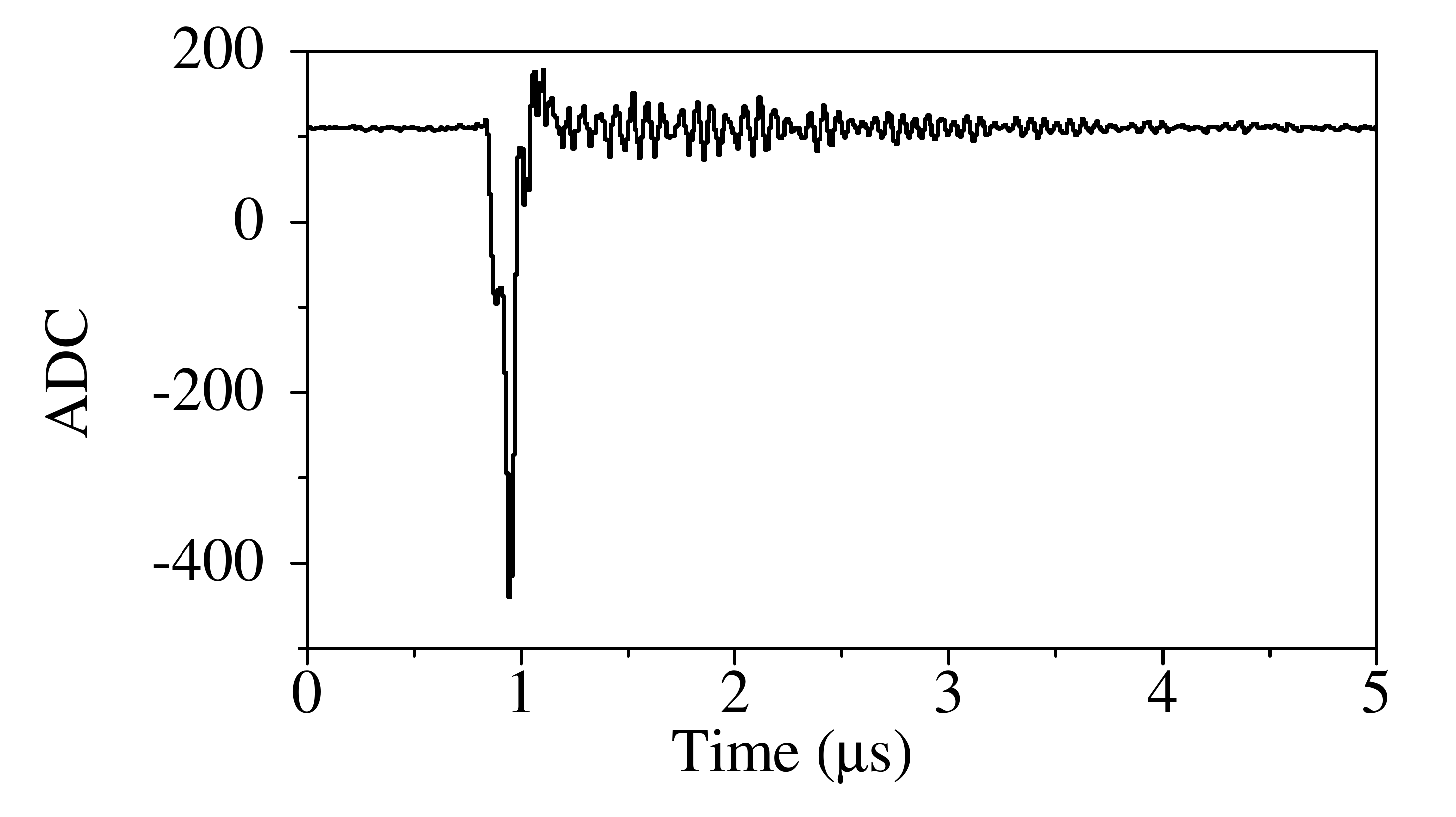}
\centering \caption{\it Example of micro-discharge event found at Module~1.}
\label{fig:M1md}
\end {figure}

Furthermore, micro-discharge rates have been studied in the prototype module and Module~1 during dedicated runs with an opposite polarity trigger. In the prototype module, there were 8 operative detectors and a dedicated run time of 28.9~h looking for micro-discharge events. Four micro-discharge events were found in four different detectors which corresponds to a rate of 0.017\,$\pm$\,0.009\,$\mu$d/h per cable. During Module~1 commissioning, another dedicated run of 18~h took place. At this phase of the commissioning 21 out of 29 Module~1 detectors were operational. In this case, 114 micro-discharge events were observed randomly distributed among detectors, corresponding to a rate of 0.30\,$\pm$\,0.03\,$\mu$d/h per cable. An example of a micro-discharge event found at Module~1 is shown in Figure~\ref{fig:M1md}. The results are summarized in Table~\ref{tab:module}. In both cases the micro-discharge rate is below the upper limits obtained in the characterization measurements shown in Section~\ref{sec4}. This is consistent with the fact that most of the detectors operate at voltages less than 5~kV, so a lower rate is expected. The rate in the prototype module is probably lower due to the lower detector operating voltages and the fact that those data were taken after allowing time for conditioning effects, reducing the effective micro-discharge rate.

\begin{table}[ht]
\begin{center}
\fboxrule=0cm \fboxsep=0cm \fbox{
\begin{tabular}[]{lll}
\toprule
Module & Detectors & Average $\mu$d rate\\
\cmidrule{1-3}															
Prototype Module & 8 & 0.017\,$\pm$\,0.009\,$\mu$d/h  \\
Module 1	& 21 &   0.30\,$\pm$\,0.03\,$\mu$d/h    \\
\bottomrule
\end{tabular}
}\caption{\it Prototype Module and Module 1 average micro-discharge rates. }
\label{tab:module}
\end{center}
\end{table}

\section{Towards a large-scale $^{76}$Ge experiment}
\label{sec6}

The results reported here can inform the design of HV component testing apparatus for a large scale Ge experiment. In a possible next generation experiment, more cables will have to be integrated in a small space. A reliable configuration that does not require all the cables and feedthroughs to be characterized must be found to save time, due to the increased number of cables and feedthroughs. Axon' cables together with the layout given in Section~\ref{sec3.1} are a plausible option as none of the cables were defective or showed a high micro-discharge rate. Regarding the feedthroughs, an improved design with more connections per flange would be desirable. In addition, more reliable feedthroughs that are all capable of holding 5~kV will be required.


\section*{Conclusions}

The phenomenon of micro-discharge induced by HV has been studied in the context of the \mj\ \dem. Initial R\&D measurements, carried out in a setup closely reproducing the cable path, led to the optimum cable configuration. We present several conclusions: the micro-discharge rate increases with the applied voltage; the coiling radius should be at least 10 times greater than the cable radius to avoid breakdown; the strip-back length of the cable should be greater than 0.5~inches; and electropolishing the copper pieces that have sharp edges reduces the micro-discharge rate.

The original ceramic pin flange design showed current fluctuations at 5\,kV, so three different solutions were studied in parallel: a flange with SHV connectors, potting the ceramic pin flange design, and a flange with new ``Pee-Wee" connectors. Two flanges of each model were tested, and all three flange models were certified for use. For technical reasons, the Pee-Wee feedthrough flange model was selected to be used at the~\mj~\dem. As a parallel result, it was found that electron emission from the gauge used in the vacuum system was able to create micro-discharge events. This effect was avoided by placing a baffle near the pressure gauge that prevents the electrons from reaching the parts being tested. Every cable and feedthrough to be used in the~\mj~\dem\ was then characterized. Leakage current testing of the feedthroughs showed that 97\% of the pins exhibited typical leakage current. The micro-discharge rate per cable was studied and no issues were found with any of the cables. Finally, the micro-discharge occurrence in the \mj\ \dem\ was studied. The micro-discharge rate is below the upper limits obtained in the characterization measurements.

\section*{Acknowledgments}

This material is based upon work supported by the U.S. Department of Energy, Office of Science, Office of Nuclear Physics under Award  Numbers DE-AC02-05CH11231, DE-AC52-06NA25396, DE-FG02-97ER41041, DE-FG02-97ER41033, DE-FG02-97ER41042, DE-SC0012612, DE-FG02-10ER41715, DE-SC0010254, and DE-FG02-97ER41020. We acknowledge support from the Particle Astrophysics Program and Nuclear Physics Program of the National Science Foundation through grant numbers PHY-0919270, PHY-1003940, 0855314, PHY-1202950, MRI 0923142 and 1003399. We acknowledge support from the Russian Foundation for Basic Research, grant No. 15-02-02919. We  acknowledge the support of the U.S. Department of Energy through the LANL/LDRD Program. This research used resources of the Oak Ridge Leadership Computing Facility, which is a DOE Office of Science User Facility supported under Contract DE-AC05-00OR22725. This research used resources of the National Energy Research Scientific Computing Center, a DOE Office of Science User Facility supported under Contract No. DE-AC02-05CH11231. We thank our hosts and colleagues at the Sanford Underground Research Facility for their support.

\nocite{*}
\bibliographystyle{elsarticle-num}
\bibliography{HVtesting}

\begin{thebibliography}{10}
\expandafter\ifx\csname url\endcsname\relax
  \def\url#1{\texttt{#1}}\fi
\expandafter\ifx\csname urlprefix\endcsname\relax\def\urlprefix{URL }\fi
\expandafter\ifx\csname href\endcsname\relax
  \def\href#1#2{#2} \def\path#1{#1}\fi

\bibitem{Zralek}
M.~Zralek, {On the Possibilities of Distinguishing Dirac from Majorana
  Neutrinos}, ACTA Phys. Pol. B 28 (1997) 2225.

\bibitem{Camilleri}
{L. Camilleri, E. Lisi and J.F. Wilkerson}, {Neutrino Masses and Mixings:
  Status and Prospects}, Ann. Rev. Nucl. Part. Sci. 58 (2008) 343.

\bibitem{Avignone}
{F. T. III Avignone, S. R. Elliott and J. Engel}, {Double beta decay, Majorana
  neutrinos, and neutrino mass}, Rev. mod. Phys. 80 (2008) 481.

\bibitem{vergados}
{J. D. Vergados, H. Ejiri and F. Simkovic}, Theory of neutrinoless double-beta
  decay, Rep. Prog. Phys. 75 (2012) 1063013.

\bibitem{mjd}
N.~Abgrall, et~al., {The \textsc{Majorana Demonstrator} Neutrinoless
  Double-Beta Decay Experiment}, Adv. High Energy Phys. 2014 (2014) 365432.

\bibitem{ppc}
P.~S. Barbeau, et~al., {Large-mass ultralow noise germanium detectors:
  performance and applications in neutrino and astroparticle physics}, JCAP 9
  (2007) 009.

\bibitem{ppc2}
P.~N. Luke, et~al., {Low capacitance large volume shaped-field germanium
  detector}, IEEE Transactions on Nuclear Science 36 (1989) 926.

\bibitem{surf}
J.~Heise, {The Sanford Underground Research Facility at Homestake}, J. Phys.:
  Conf. Ser. 606 (2015) 012015.

\bibitem{mjdbkg}
C.~{Cuesta}, et~al., {Background Model for the \textsc{Majorana Demonstrator}},
  Phys.~Proc. 61 (2015) 821.

\bibitem{MJDAssay}
N.~{Abgrall}, et~al., {The \textsc{Majorana} Radioassay Program}, arXiv
  1601.03779.

\bibitem{md}
K.~Heeger, et~al., {High-Voltage Micro Discharge in Ultra-Low Background
  $^{3}$He Proportional Counters}, IEEE 47 (2000) 1829.

\bibitem{md2}
T.~Ono, et~al., {Micro-discharge and electric breakdown in a micro-gap}, J.
  Micromech. Microeng. 10 (2000) 445.

\bibitem{mdATLAS}
T.~Kuwano, et~al., {Systematic study of micro-discharge characteristics of
  ATLAS barrel silicon microstrip modules}, Nucl. Inst. and Meth. A 579 (2007)
  782.

\bibitem{SNO}
J.~Amsbaugh, et~al., {An array of low-background $^{3}$He proportional counters
  for the Sudbury Neutrino Observatory}, Nucl. Inst. and Meth. A 579 (2007)
  1054.

\bibitem{vacuum}
P.~Redhead, {Measurement of vacuum; 1950–2003}, J. of Vacuum Science \& Tech.
  A 21 (2003) S1.

\end{thebibliography}







\end{document}